\documentclass[aps,prl,reprint,groupedaddress,showpacs]{revtex4-1}

\usepackage{amsmath}
\usepackage{amssymb}
\usepackage{graphicx}

\newcommand{\BF}{\mathrm{BF}}
\newcommand{\BB}{\mathrm{BB}}
\newcommand{\B}{\mathrm{B}}
\newcommand{\F}{\mathrm{F}}

\begin{document}


\title{Coherent Interaction of a Single Fermion with a Small Bosonic Field}



\author{Sebastian Will}
\email{sebastian.will@lmu.de}
\author{Thorsten Best}
\altaffiliation{Now: Albert-Ludwigs-Universit\"at, 79104 Freiburg, Germany}
\author{Simon Braun}
\author{Ulrich Schneider}
\author{Immanuel Bloch}
\affiliation{Institut f\"ur Physik, Johannes Gutenberg-Universit\"at, 55099 Mainz, Germany}
\affiliation{Ludwig-Maximilians-Universit\"at, 80799 M\"unchen, Germany}
\affiliation{Max-Planck-Institut f\"ur Quantenoptik, 85748 Garching, Germany}

\date{\today}

\begin{abstract}
We have experimentally studied few-body impurity systems consisting of a single fermionic atom and a small bosonic field on the sites of an optical lattice. Quantum phase revival spectroscopy has allowed us to accurately measure the absolute strength of Bose-Fermi interactions as a function of the interspecies scattering length. Furthermore, we observe the modification of Bose-Bose interactions that is induced by the interacting fermion. Due to an interference between Bose-Bose and Bose-Fermi phase dynamics, we can infer the mean fermionic filling of the mixture and quantify its increase (decrease) when the lattice is loaded with attractive (repulsive) interspecies interactions. 
\end{abstract}

\pacs{03.75.Lm, 67.85.Pq, 21.45.-v}
\keywords{}
\maketitle

Multi-component systems play a central role in quantum many-body physics. From interacting atoms and photons to electrons and phonons, the interplay of interactions in binary mixtures gives rise to intriguing quantum phenomena such as superradiance, BCS superfluidity or polaron physics \cite{Alexandrov:1994, *Bruderer:2007,*Tempere:2009, *Privitera:2010}. Recently, especially the problem of an impurity embedded in an external quantum environment has been addressed experimentally, such as fermionic spin impurities in a Fermi sea, which lead to the observation of a Fermi polaron \cite{Schirotzek:2009,*Nascimbene:2009}, or an ion immersed in a Bose-Einstein condensate \cite{Zipkes:2010,*Schmid:2010}. When such impurity systems are scaled down to the few-body regime, they can share important properties with models for atomic nuclei \cite{Platter:2009}.

Here we present the experimental study of an elementary few-particle system consisting of a single fermion and a coherent bosonic field. While research on Bose-Fermi mixtures in optical lattices has so far mainly focussed on the coherence properties of the global quantum many-body state \cite{Albus:2003, Guenter:2006, *Ospelkaus:2006a,Best:2009}, we investigate the local properties of miniature Bose-Fermi systems on individual lattice sites. Using quantum phase revival spectroscopy \cite{Greiner:2002b, *Sebby:2007, Will:2010}, we precisely measure the absolute strength of intra- and interspecies interactions as a function of the interspecies scattering length, tuned by means of a Fesh\-bach resonance. Already moderate Bose-Fermi interactions give rise to notable changes of the on-site wavefunctions, which we observe as modifications of the Bose-Bose interaction  \cite{Alon:2005, Luehmann:2008, *Johnson:2009, *Buechler:2010,*Dutta:2010,*Mering:2010}. Furthermore, we have devised a novel method that allows us to selectively infer the mean fermionic filling within the volume, in which bosons and fermions overlap.

\begin{figure}
\includegraphics[width=1.0\columnwidth]{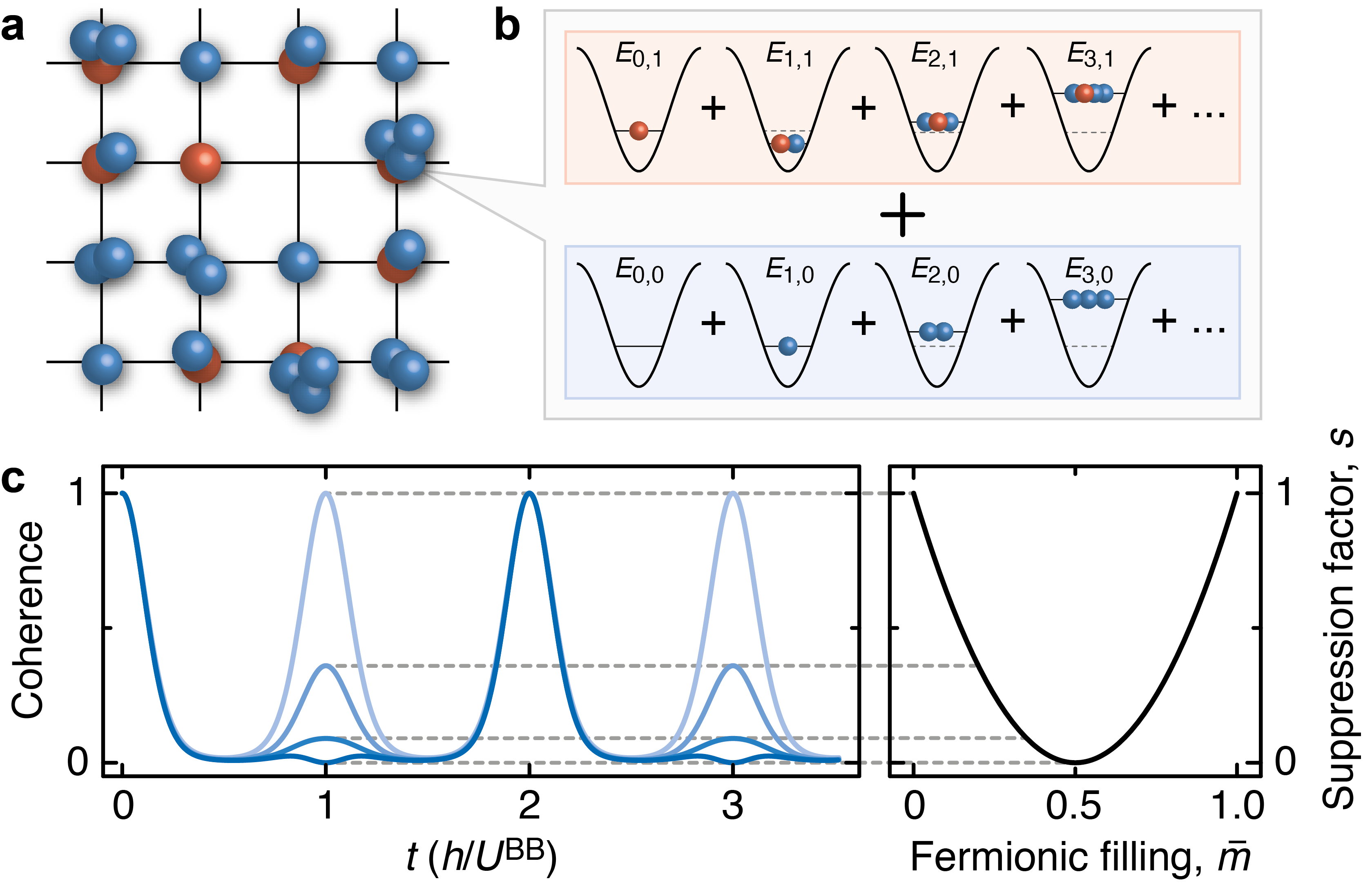}
\caption{\label{fig:cartoon} Quantum phase revivals in a few-body Bose-Fermi system. (a) In a shallow optical lattice both the bosonic and the fermionic species are delocalized and show atom number fluctuations. The cartoon picture displays a possible result of a projection measurement assuming a mean fermionic filling of $\bar{m}=0.5$. Blue (red) balls indicate bosons (fermions). (b) The local quantum state at a lattice site corresponds to coherent superpositions of bosonic atom number states \emph{with} and \emph{without} the presence of a fermion with eigenenergies $E_{n,m}$ (see text).  (c) Suppression of odd quantum phase revivals in a Bose-Fermi mixture at $|U^\BF /U^\BB| = 0.5$ for variable fermionic fillings: $\bar{m} = 0$, $0.2$, $0.35$ and $0.5$ (blue solid lines, darker color for larger $\bar{m}$). The right panel displays the suppression factor \emph{s} as a function of the filling $\bar{m}$. 
}
\end{figure}

We start with a delocalized Bose-Fermi mixture in a shallow optical lattice. Then we rapidly increase the lattice depth, suppressing both bosonic and fermionic tunneling ($J_B, J_F \rightarrow 0$) and freezing out the delocalized atom distributions of bosons and fermions. In this setting, the eigenstates at a lattice site are given by product atom number states $|n\rangle | m\rangle$, containing  $n$ bosons (where $n$ is a non-negative integer) and $m$ fermions (where $m$ is either 0 or 1); the respective eigenenergies are denoted by $E_{n,m}$ (Fig.~\ref{fig:cartoon}). For a delocalized mixture, both the bosonic and fermionic component show number fluctuations and the corresponding on-site quantum states can be described as a coherent superposition of bosonic atom number states both in the absence ($m=0$) and the presence ($m=1$) of a fermion. The resulting time evolution at a lattice site can be modeled as a superposition of phase evolutions with and without a fermion, which are governed by the eigenenergies $E_{n,0}$ and $E_{n,1}$,
\begin{equation}
\label{eq:wavefunction}
|\psi_{\mathrm{BF}} (t) \rangle = \sum_{n=0}^{\infty} c_{n} e^{-i E_{n,0} t/ \hbar} |n \rangle |0 \rangle \,+\,  d_{n} e^{-i E_{n,1} t/ \hbar} |n \rangle |1 \rangle. \nonumber
\end{equation}
Here $c_n$ and $d_n$ denote the probability amplitudes of finding $n$ bosons without ($m=0$) and with ($m=1$) a fermion, respectively. 

Within the single-band Bose-Fermi Hubbard model \cite{Albus:2003} the eigenenergies are given by 
\begin{equation}
\label{eq:eigenenergies}
E_{n,m}=\frac{U^{\mathrm{BB}}}{2} \cdot n  (n-1)  +  U^{\mathrm{BF}} \cdot n \, m.
\end{equation} 
Here the Bose-Bose interaction energy, $U^{\mathrm{BB}} \propto a_{\mathrm{BB}} \int |\phi_\B (\mathbf{r})|^4 d^3 r$, 
and the Bose-Fermi interaction energy,   
$ U^{\mathrm{BF}} \propto  a_{\mathrm{BF}} \int |\phi_\B (\mathbf{r})|^2 |\phi_\F  (\mathbf{r})|^2 d^3 r$,
are independent of the bosonic and fermionic atom numbers $n$ and $m$, since the model is restricted to the lowest lattice band; $a_{\mathrm{BB}}$ ($a_{\mathrm{BF}}$) denotes the intraspecies (interspecies) scattering length and  $\phi_{\mathrm{B}}(\mathbf{r})$ ($\phi_{\mathrm{F}}(\mathbf{r})$) the bosonic (fermionic) ground state orbital at a lattice site. While essential features of the resulting quantum dynamics are captured by this single-orbital model, the observed dynamics contain signatures that can only be explained within a multi-orbital approach. Here, the interaction-induced deformation of on-site wavefunctions gives rise to interaction strengths $U_{n,m}^\BB$ and $U_{n}^\BF$ that explicitly depend on the number of bosons and fermions \cite{Luehmann:2008, Dutta:2010}. 


\begin{figure}
\includegraphics[width=0.9\columnwidth]{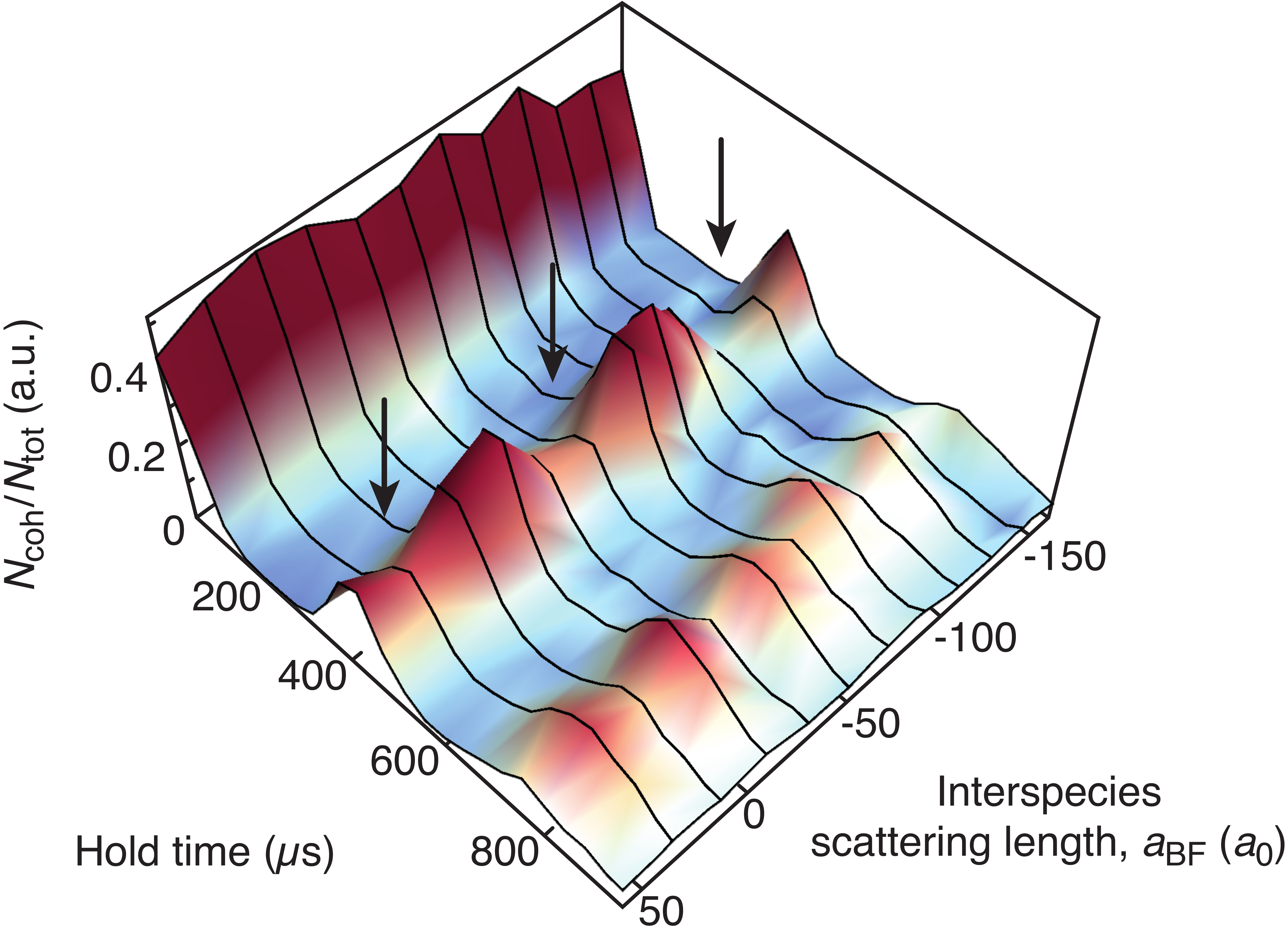}%
\caption{\label{fig:peakosc} Interference of Bose-Bose and Bose-Fermi phase dynamics. The initial quantum phase revivals are modulated as a function of the interspecies scattering length $a_\BF$; the intraspecies scattering length $a_\BB$ is fixed at $+102(2)$$a_0$ \cite{Will:2010}. Suppressions of the first revival happen at $a_{\mathrm{BF}} = -126(2)$$a_0$, $-40(3)$$a_0$ and $+41(3)$$a_0$, as marked by the arrows. Black lines indicate the traces recorded in the experiment.
}
\end{figure}


The interactions between the bosonic field and the fermion are encoded in the dynamics of the on-site wavefunction, $| \psi_\BF (t) \rangle$. Quantum phase revival spectroscopy \cite{Will:2010} allows us to probe the corresponding interaction energies by sampling the time-dependent coherence of the bosonic component (see below). It is proportional to $|\langle \psi_\BF(t) | \hat{a} |\psi_\BF(t) \rangle |^2 \equiv |\langle \hat{a} \rangle (t) |^2$, where $\hat{a}$ denotes the annihilation operator for a bosonic atom at a lattice site. The time evolution of the coherence is governed by the interference of the two dynamical evolutions with and without a fermion: 
\begin{equation}
|\langle \hat{a} \rangle (t) |^2 = \left|\sum_{n=0}^{\infty} C_n(t) + D_n(t)\right|^2,
\end{equation}
where the purely bosonic contribution enters as
\begin{equation}
C_n(t)  = \sqrt{n+1} \, c_n^* c_{n+1} \,e^{-i(E_{n+1,0} - E_{n,0})t/\hbar} \nonumber 
\end{equation}
and the Bose-Fermi interactions are contained in the term
\begin{equation}
D_n(t)  = \sqrt{n+1} \, d_n^* d_{n+1} \,e^{-i(E_{n+1,1} - E_{n,1})t/\hbar}. \nonumber
\end{equation}

Generally, the resulting collapse and revival dynamics of the coherence are rather complex. However, an analytic expression can be derived when the influence of the Bose-Bose and the Bose-Fermi interaction on the bosonic atom number statistics is neglected, which is justified in the limit of small interactions. This motivates the use of coherent states for the bosonic field, whose probability amplitudes scaled by the mean fermionic atom number $\bar{m}$, read $c_n = \sqrt{1- \bar{m}}\,e^{-|\alpha|^2/2 } \alpha^n/\sqrt{n!}$ and  $d_n = \sqrt{\bar{m}}\,e^{-|\alpha|^2/2 } \alpha^n/\sqrt{n!}$, where $\alpha= \sqrt{\bar{n}} e^{i\phi}$ denotes the complex field amplitude with mean bosonic atom number $\bar{n}$ and initial phase $\phi$.  With the additional assumption of single-orbital eigenenergies according to Eq.~(\ref{eq:eigenenergies}), we obtain the quantum phase evolution
\begin{eqnarray}
\label{eq:idealdyn}
& &|\langle \hat{a} \rangle (t) |^2/\bar{n} = e^{2 \bar{n}(\cos(U^\BB t/\hbar)-1)} \nonumber \\
& &\quad \quad \quad \times \left\{ 1 - 2  \bar{m} (1-  \bar{m})\left[1 - \cos \left(U^\BF t/ \hbar \right)\right] \right\}.
\end{eqnarray}

These idealized dynamics are illustrated in Fig.~\ref{fig:cartoon}(c) for several mean fermionic fillings $\bar{m}$. The suppression of revivals is a striking signature for the interference between the Bose-Bose and the Bose-Fermi phase evolution. Particularly, for the case $U^\BF /U^\BB=z + 0.5$ $(z \in \mathbb{Z})$ the suppression factor has a simple relation to the mean fermionic filling, $s=(1- 2 \bar{m})^2$ [Fig.~\ref{fig:cartoon}(c), right panel].  

The experiment starts with the preparation of a quantum degenerate mixture of $1.7(3)\times 10^{5}$ bosonic $^{87}$Rb and $2.1(4)\times 10^{5}$ fermionic $^{40}$K atoms in their respective hyperfine ground states $|1, +1\rangle$ and $|9/2, -9/2 \rangle$. Using the interspecies Feshbach resonance at \mbox{$546.75(6)$ G} \cite{Simoni:2008}, we tune $a_{\mathrm{BF}}$ between $-161.2(1)$$a_0$ and $+134(19)$$a_0$, where $a_0$ is the Bohr radius. Then, a three-dimensional (3D) optical lattice ($\lambda=738$ nm) is adiabatically ramped up to a depth of $V_{\mathrm{L}}=5.2$ $E_{\mathrm{rec}}^{\mathrm{B}}$, where $E_{\mathrm{rec}}^{\mathrm{B}}=h^2/(2 m_{\mathrm{B}} \lambda^2)$ is the recoil energy for $^{87}$Rb. At this lattice depth and for all interspecies interactions used in our measurements, we expect the bosons to form a superfluid and the fermions to be delocalized \cite{Best:2009}. A 3D array of coherent bosonic fields with partial fermionic filling is created by rapidly increasing the lattice depth from $V_\mathrm{L}$ to $V_\mathrm{H} = 28.2(3)$ $E_{\mathrm{rec}}^{\mathrm{B}}$. This suppresses the tunnel coupling, freezes out the atom distributions and initiates the quantum phase evolution at each lattice site. After variable hold times $t$, all trapping potentials are switched off, the bosonic and fermionic clouds expand during 10 ms time-of-flight and an absorption image of the bosonic interference pattern is recorded. As a measure of the bosonic coherence $|\langle \hat{a} \rangle (t) |^2$, we evaluate the ratio of the summed atom numbers in the central, first- and second-order coherence peaks to the total atom number, $N_{\mathrm{coh} }/ N_{\mathrm{tot}}$  \cite{Greiner:2002b}.

The experimental data reveals a modulation of the initial quantum phase revivals depending on the interspecies interaction strength (Fig.~\ref{fig:peakosc}). For the first revival we detect three local minima with suppression factors $s=0.57(3)$, $0.43(3)$ and $0.16(2)$ as the attraction is increased. From this we derive the mean fermionic fillings of the mixture to be $\bar{m}=0.12(1)$, $0.17(1)$ and $0.30(1)$, respectively, according to $\bar{m} = (1 \mp \sqrt{s})/2$, where the minus (plus) sign corresponds to $\bar{m} < 0.5$ ($\bar{m} > 0.5$). Here the minus sign is chosen, since fermionic fillings $\bar{m}>0.5$ are not expected for our experimental parameters. Qualitatively, the dynamics are well captured by the single-orbital coherent state model of Eq.~(\ref{eq:idealdyn}), as multi-orbital effects play a minor role for short observation times.


\begin{figure}
\includegraphics[width=0.92\columnwidth]{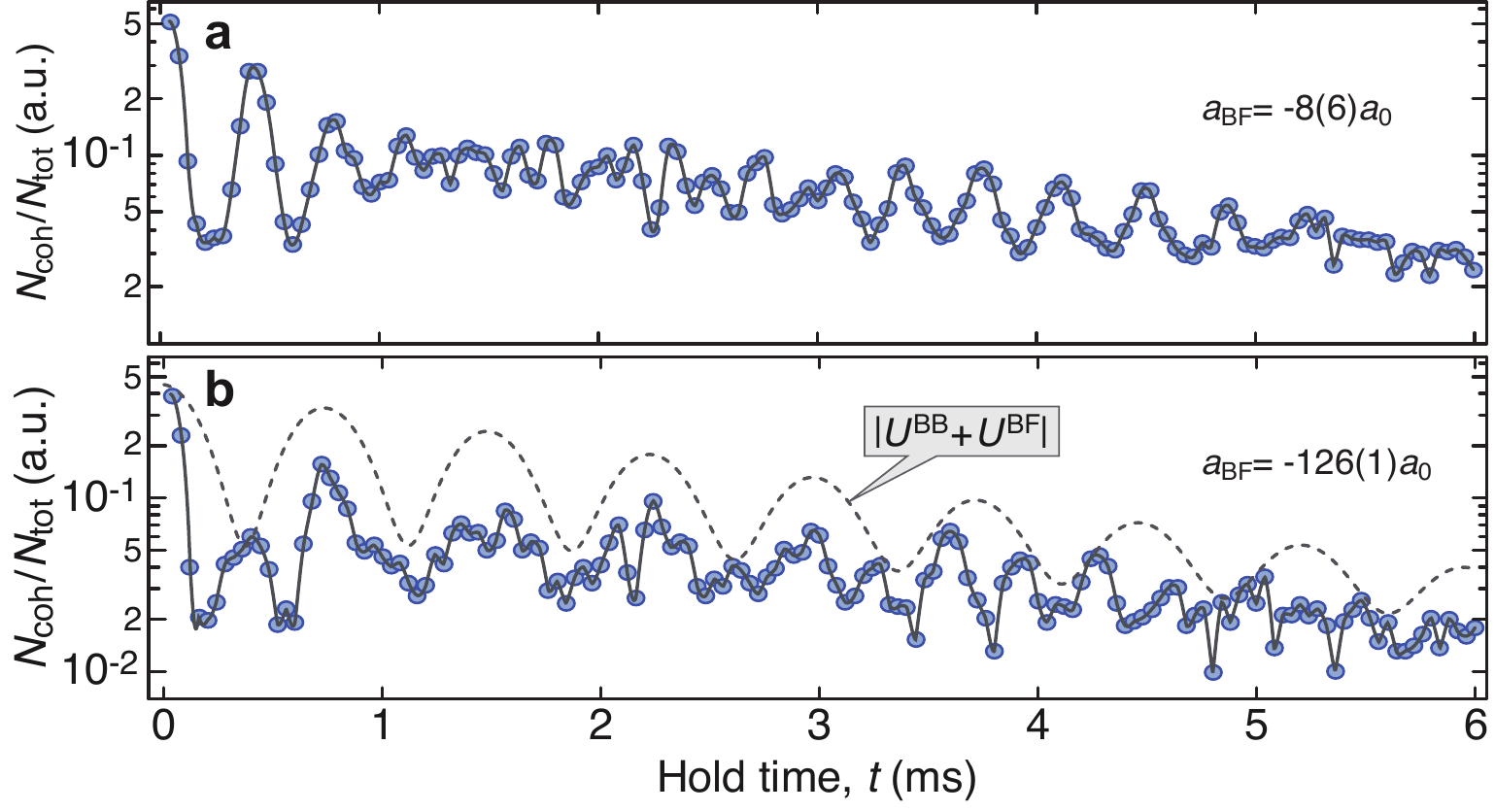}
\caption{\label{fig:traces} Quantum phase revival traces of the Bose-Fermi system for vanishing, (a), and strong interspecies attraction, (b). In the later case, every second revival is suppressed by an envelope (gray dashed line), which corresponds to the spectral components of order $|U^\BB + U^\BF|$. Each data point represents a single run of the experiment. The solid lines interpolate the data and serve as a guide to the eye.}
\end{figure}



\begin{figure}
\includegraphics[width=1.0\columnwidth]{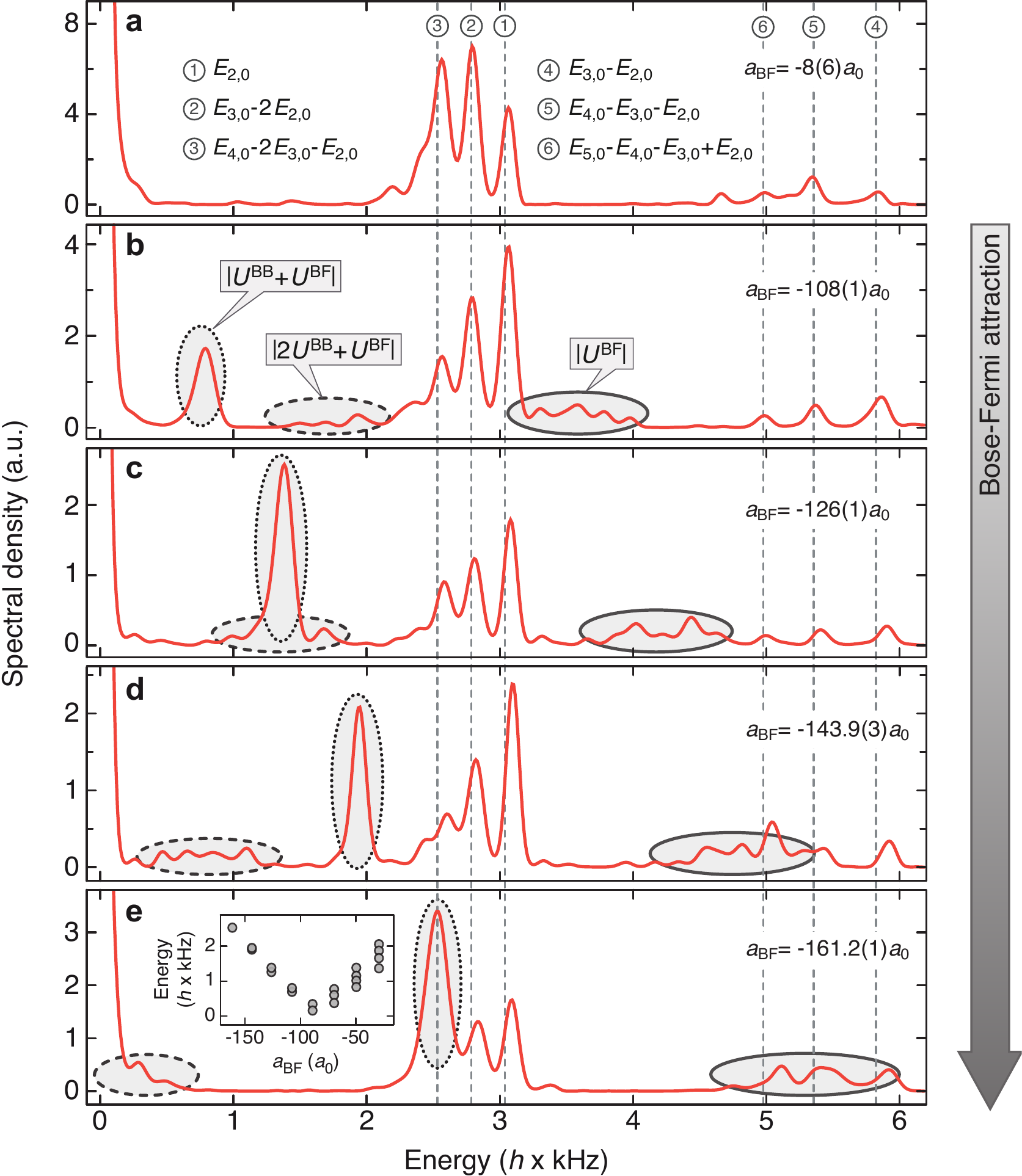}
\caption{\label{fig:spectra} Fourier spectra of the time traces for vanishing, (a), and linearly increasing Bose-Fermi attraction, (b) to (e). The shaded ovals highlight the contributions involving direct Bose-Fermi interactions. The inset of panel (e) shows the \mbox{$|U^\BB + U^\BF|$} components as a function of $a_\BF$. Dashed vertical lines indicate the spectral contributions of orders $U^\BB$ ({\textcircled{\scriptsize 1}} to {\textcircled{\scriptsize 3}}) and $2 U^\BB$ \mbox{({\textcircled{\scriptsize 4}} to {\textcircled{\scriptsize 6}})} for a purely bosonic system.}
\end{figure}


Quantitative information about the interactions in the Bose-Fermi system is obtained by sampling long time traces of quantum phase revivals yielding high spectral resolution (Fig.~\ref{fig:traces}). The trace at vanishing interspecies attraction and its Fourier spectrum [Figs.~\ref{fig:traces}(a) and \ref{fig:spectra}(a)] show the signatures of effective multi-body interactions as previously observed in a purely bosonic system \cite{Johnson:2009,Will:2010}. However, at stronger attraction new striking features appear in the spectra: First, prominent additional peaks arising from direct Bose-Fermi interactions and, second, a small, but significant upshift of the Bose-Bose interaction energies at orders $U^\BB$ and $2 U^\BB$ (Fig.~\ref{fig:spectra}).  

The emerging Bose-Fermi features of orders $|U^\BF|$, $|U^\BB+ U^\BF|$ and $|2 U^\BB+ U^\BF|$  exhibit an almost linear dependence on $a_\BF$ [Fig.~\ref{fig:spectra}(b) to (e)]. Generally, each of them consists of a comb of energies owing to the explicit boson number dependence of $U_{n,m}^\BB$ and $U_{n}^\BF$; e.g.~for local bosonic occupations of up to six atoms (which corresponds to the appearance of four peaks of order $U^\BB$), up to six components of order $|U^\BF|$, five of order $|U^\BB+ U^\BF|$ and four of order $|2 U^\BB+ U^\BF|$ are  expected. However, the separation between the individual components varies as a function of $a_\BF$; for example the energies of order $|U^\BB+ U^\BF|$ appear to narrow down to a single peak for $-150$$a_0 \lesssim a_\BF \lesssim -100$$a_0$ [see inset of Fig.~\ref{fig:spectra}(e)], because the interaction-induced changes of $U_{n,m}^\BB$ and $U_{n}^\BF$ approximately compensate each other due to opposite signs.


\begin{figure}
\includegraphics[width=1.0\columnwidth]{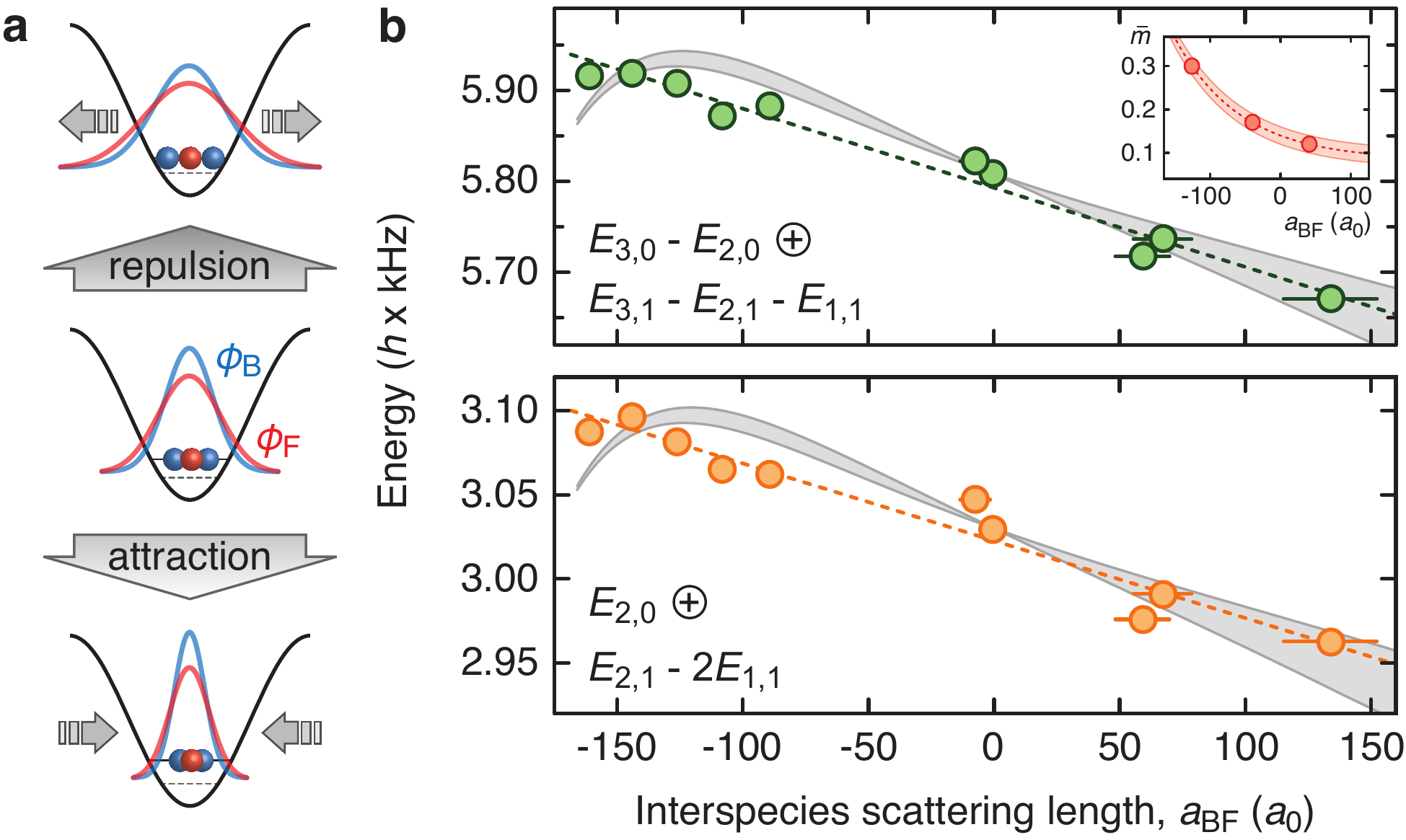}%
\caption{
\label{fig:ubbshift} Modification of the Bose-Bose interaction strength induced by an interacting fermion. (a) Repulsive (attractive) interspecies interactions broaden (shrink) the on-site orbitals $\phi_\B(\mathbf{r})$ and $\phi_\F(\mathbf{r})$ and thereby affect the Bose-Bose interaction strengths $U^\BB_{n,1}$. (b)  The highest spectral components of order $U^\BB$ ($2 U^\BB$) are shown as a function of $a_\BF$ in the lower (upper) panel. The underlying energies $E_{2,0}$ and $E_{2,1} - 2 E_{1,1}$ ($E_{3,0}-E_{2,0}$ and $E_{3,1}-E_{2,1} -E_{1,1}$) cannot be resolved individually and appear as a single superposition peak in the spectra. The dashed line is a linear fit with a slope of  $-0.46(4)$ Hz/$a_{\mathrm{0}}$ ($-0.87(5)$  Hz/$a_{\mathrm{0}}$). Shaded areas show the shift calculated within a variational model, which takes into account the measured values for the fermionic filling fitted by an exponential function, $\bar{m}(a_\BF)$ (inset). The data at $a_\BF = 0$$a_0$ have been obtained in a sample without fermions.}
\end{figure}


The observed shift of the Bose-Bose interaction strengths is induced by the presence of an interacting fermion in analogy to the atom number dependence of $U^\BB$ in a purely bosonic system \cite{Will:2010}. For the case of repulsive interspecies interactions a broadening of the on-site wavefunctions $\phi_\B(\mathbf{r})$ and $\phi_\F(\mathbf{r})$ is expected and the Bose-Bose interaction strength is effectively reduced; the reverse happens for attractive interspecies interactions [Fig.~\ref{fig:ubbshift}(a)]. In Fig.~\ref{fig:ubbshift}(b) these modifications are shown for two selected components of order $U^\BB$ and $2U^\BB$ for both negative and positive $a_\BF$. Using the linear slopes extracted from the data and assuming a maximal fermionic filling $\bar{m} \approx 0.3$, we derive conservative upper bounds for the shifts $(\partial U_{2,1}^\BB/\partial a_\BF)/h <-1.5(2)$ Hz/$a_0$ and $(\partial U_{3,1}^\BB/\partial a_\BF)/h <-1.4(2)$ Hz/$a_0$. With a variational harmonic oscillator model we have confirmed that the observed shifts have the correct order of magnitude. However, the good agreement must be regarded fortuitous. These findings quantitatively support earlier, indirect evidence for a renormalization of Hubbard parameters due to interspecies interactions \cite{Best:2009, Luehmann:2008}.

In conclusion, we have studied the interactions in elementary few-body systems formed by a single fermion and a small bosonic field at the sites of an optical lattice. Using quantum phase revival spectroscopy, we have been able to accurately measure the mean fermionic filling and the Bose-Fermi interaction strength as a function of the interspecies scattering length. While in the latter case an essentially linear dependence on the interspecies scattering length is found (Fig.~\ref{fig:spectra}), we additionally observe how the interaction among the bosons is modified, mediated via Bose-Fermi interactions (Fig.~\ref{fig:ubbshift}). 

Miniature impurity systems, as the one presented here, are suited to investigate polaron physics in ultracold quantum gases \cite{Alexandrov:1994,Bruderer:2007, Tempere:2009, Privitera:2010} and form an ideal test bed for effective field theories \cite{Johnson:2009}, that are highly relevant to the description of atomic nuclei \cite{Platter:2009}. Our measurement technique might enable thermometry in Bose-Fermi mixtures based on a temperature dependence of the fermionic filling and allows for exact absolute measurements of two- and higher-body interaction energies \cite{Luehmann:2008, Johnson:2009, Buechler:2010,Dutta:2010} in a multi-component quantum system. Furthermore, the control of interspecies collisions demonstrated here is likely to play an important role in the realization of advanced quantum information processing schemes, which rely on the use of different atomic species for data storage and gate operation \cite{Albus:2003,You:2000,*Calarco:2004,*Soderberg:2009}. 

\begin{acknowledgments}
We thank D.-S.~L\"uhmann for inspiring discussions. This work was supported by the DFG (FOR801), the EU (SCALA), the US ARO with funding from DARPA (OLE program), the AFOSR, MATCOR (S.W.) and the Gutenberg-Akademie (S.W.)
\end{acknowledgments}

%

\end{document}